# Modeling the spatial dynamics of income in cities

Vincent Verbavatz[1,2], Marc Barthelemy[1,3,*]

**Affiliations:**

[1]Institut de Physique Théorique, CEA, CNRS-URA 2306, F-91191, Gif-sur-Yvette, France.

[2]École Nationale des Ponts et Chaussées, Champs-sur-Marne, France.

[3]Centre d'Analyse et de Mathématique Sociales, CNRS/EHESS, 54 Boulevard Raspail, 75006 Paris, France.

*Corresponding author. Email: marc.barthelemy@ipht.fr

**Abstract:**

Urban inequality is a major challenge for cities in the 21st century. This inequality is reflected in the spatial income structure of cities which evolves in time through various processes. Gentrification is a well-known illustration of these dynamics in which the population of a low income area changes as wealthier residents arrive and old-settled residents are expelled. Less understood but very important is the reverse process of gentrification through which areas of cities get impoverished. Gentrification has been widely studied among social sciences, especially in case studies, but there have been fewer quantitative analyses of this phenomenon, and more generally about the spatial dynamics of income in cities. Here, we first propose a quantitative analysis of these income dynamics in cities based on household incomes in 45 American and 9 French Functional Urban Areas (FUA). We found that an important ingredient that determines the evolution of the income level of an area is the income level of its immediate neighboring areas. This empirical finding leads to the idea that these dynamics can be modeled by the voter model of statistical physics. We show that such a model constitutes an interesting tool for both describing and predicting evolution scenarios of urban areas with a very limited number of parameters (two for the US and one for France). We illustrate our results by computing the probability that areas will change their income status in the case of Boston and Paris at the horizon of 2030.

## Introduction

Urban inequality is a crucial topic in social studies (Nijman & Wei, 2020) and represents a primary challenge for the development of cities. In particular, the spatial dynamics of inequality are often seen through the lens of gentrification. This phenomenon is the multistage process through which low-income areas transform with the influx of more affluent residents (Lees, et al., 2013; Lees, et al., 2010). As the new population settles in, the economic value of the area increases, new businesses develop and real estate prices go up while long-established populations eventually have to move out due to the rise of rents (Freeman & Braconi, 2004; Atkinson, et al., 2011). This process goes along gentrification which has been widely studied over the past 40 years (Smith, 1979) for specific cases (Freeman & Braconi, 2004; Fujitsuka, 2005; Clerval, 2008; Torrens & Nara, 2007; Pattaroni, et al., 2012; Venerandi, et al., 2017; Doring & Ulbricht, 2018), or about the role of public investment (Zuk, et al., 2018) and its impact on health for example (Gibbons, et al., 2018). The inverse process of gentrification (Atkinson, 2004; Douglas & Massey, 1993; Lobmayer & Wilkinson, 2002), through which districts become impoverished has been, however, far less conceptualized, leading previous authors to question the relevance of the concept of gentrification *per se* (Bourdin, 2008). The two phenomena are however correlated as the outflow of people generated by gentrification generally results in other processes (urban sprawl, impoverishment) somewhere else in the city. More generally, a major question in urban studies is then to understand why and how certain districts can become richer or poorer (Rosenthal & Ross, 2015), and what governs the household demand (Ioannides & Zabel, 2008). In the following, we will reduce this complex issue which has multiple social, demographic, cultural or economic implications, to the more quantitative and tractable problem of local income dynamics, thus putting aside qualitative aspects of gentrification and impoverishment.

Most quantitative studies of income dynamics in cities (such as (Rey & Montouri, 1999), and (Torrens & Nara, 2007), (Pattaroni, et al., 2012) for example) rely on econometric tools (such as multivariate regression for example). We propose here a completely different approach, coming from statistical physics. New approaches, in particular those using statistical physics tools have proven to be very fruitful to describe social phenomena such as opinion, cultural and language dynamics, crowd behavior, humans dynamics or social spreading for example (Castellano, et al., 2009). The common focus of these studies is to understand the emergence of collective phenomena from the interactions of individuals. Such an approach was also initiated in sociology with the famous example of the Schelling model for segregation (Schelling, 1971). We propose here a model in the same spirit for the income dynamics in urban areas. This model is parsimonious, based on a simple rule and is not purely descriptive. We show that at the simplest level the probability for a specific neighborhood to change its income level is mainly governed by the income level of its neighbors, and the resulting model is known as the voter model in the statistical physics literature (Holley & Liggett, 1975; Krapivsky, et al., 2010). Although this model is very simple, it is particularly well-suited to describe and predict the local dynamics of household incomes in cities. We show that the social evolution of a district significantly depends on the economic level of its direct neighbors and that such a representation is sufficient enough to statistically predict the social dynamics of incomes in cities.

**Household income in US cities**

We use household income data at the tract level for FUAs in the US and at the IRIS level in France. This allows us to study the income level dynamics at this small spatial scale. More precisely, our analysis is based on the open datasets of median household incomes in American census tracts for each year between 2010 and 2019 (Manson, et al., 2021) (these values are obtained as interpolation between successive censuses) and median household income in French IRIS (a French equivalent of tracts) between 2001 and 2017 (INSEE, 2022) (note that for both cases, these income values are before taxes). In the US, we crossed these data with boundaries of FUAs and kept FUAs with more than 200 tracts, leading to a final set of 45 urban areas. In France, we crossed these data with boundaries of FUAs and kept FUAs with more than 150 IRIS, leading to a final set of 9 cities. In each city, we compute every year the average of the median household incomes at the tract level

$$\mu(t) = \overline{s_i(t)} \quad (1)$$

where $s_i(t)$ is the median household income of tract $i$ at time $t$ and the overbar denotes the average over all tracts. We then divide census tracts into two sets - below-average ($s_i(t) < \mu(t)$) or above-average ($s_i(t) \geq \mu(t)$) - and we define the status (or category) of tract $i$ as

$$S_i(t) = \begin{cases} -1 \text{ if } s_i(t) < \mu(t) \\ 1 \text{ if } s_i(t) \geq \mu(t) \end{cases} \quad (2)$$

Since we observe unusual one-year drops of household incomes in the data, we use a convolution regularization (see Methods) to smooth out one-year outliers. This classification hence gives a very simple city-relative definition of area wealth. Every city has tracts in both sets with a relative share of each category close to 50%. Although very basic, this definition is enough to track local evolutions of income over time and to single out significant and consistent changes.

In our datasets, we can follow the share of each group over 10 years. In particular, we note that in most American cities the share of people living in below-average tracts has decreased over 10 years (on average by 0.6%) but in more than 42% of cities it has increased (see Fig. 1), from which we conclude that there is no global and steady tendency towards an increase or a decrease of social segregation based on income-level at the US scale.

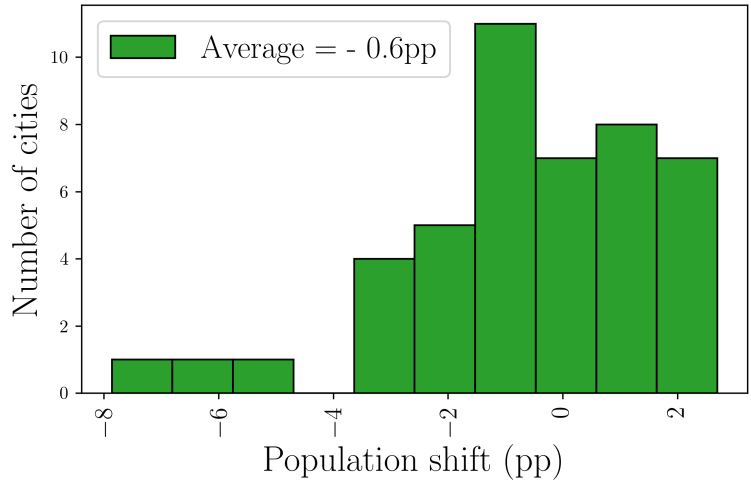

**Figure 1.** Distribution of the shift of the share of people living in tracts below average income between 2010 and 2019 in American FUAs. For example, 3 cities have experienced during this time a decrease larger than 4%. On average, the number of people living in tracts below average income has decreased from 53.6% to 53.0%. Yet, in 42% of American FUAs, this share has increased (here, "pp" stands for percentage point where a percentage point or percent point is the arithmetic difference between two percentages).

One natural question is then to understand why certain areas can become richer or poorer and not others. Instead of studying a variety of socio-economic features that could account for such changes, we focus on the impact of direct neighbors on the probability for a tract to change its category. The effect of neighbors was already considered in the gentrification context in (Guerrieri, et al., 2013). To test the effect of neighbors, we build an unweighted and undirected graph of tracts for each city. In this graph, tracts are neighbors if and only if they share a geographical border. At each time step (we tested different timesteps, see Methods), we compare $S_i(t)$ to $S_i(t-1)$ and to the share $f_i$ of neighbors that were previously in the opposite status

$$f_i = \frac{|j \in N(i) \text{ s.t. } S_j(t-1) \neq S_i(t-1)|}{|N(i)|} \quad (3)$$

where $N(i)$ is the set of the neighbors of $i$. When the status of tract $i$ changes between $t$ and $t+1$, we say the tract has flipped. The flipping probability is hence defined as the probability that a tract being at a certain income level at time $t$, will 'flip' to the other income level at time $t+1$. The assumption that we will test in the following is whether the main ingredient in the flipping probability is the share $f_i$.

A complementary question is whether the flipping is symmetric ($P(-1 \rightarrow +1) = P(+1 \rightarrow -1)$) which we test with a 4-parameter linear regression with intercept of the form

$$\text{Prob}(S_i(t) \neq S_i(t-1)) = w_0 + w_1 f_i + (w_2 + w_3 f_i)\delta_{1,S_i(t-1)} + \varepsilon \quad (4)$$

where $\delta$ is the Kronecker symbol ($\delta_{ij} = 1$ if $i = j$ and $\delta_{ij} = 0$ if $i \neq j$) and $\varepsilon$ is a centered Gaussian white noise. Non-zero values for $w_2$ or $w_3$ then implies an asymmetry of the flipping probabilities. The results of the fit are given in Table 1 considering all tracts within all cities at the same time. The $R^2$ is 0.40 in the US and 0.72 in France. The values of $w_0$ and $w_1$ are positive meaning that the probability for a tract to flip increases with the share of disagreeing neighbors, as naïvely expected. The $p$-values of coefficients $w_2$ and $w_3$ are, however, too high to reject the null hypothesis that they are irrelevant and should then be discarded. Hence, we must reject the assumption that probability to flip from -1 to 1 is different than the reverse process and conclude that flipping probability is symmetric and that on average there is no tendency to observe an increase of the average income level rather than the opposite. We thus can write (fits are shown on Fig. 2)

$$\text{Prob}(S_i(t) \neq S_i(t-1)) \approx w_0 + w_1 f_i \quad (5)$$

| Country | Coefficient | Value | $p$-value |
|---|---|---|---|
| US | $w_0$ | $0.027 \pm 0.003$ | 0 |
| | $w_1$ | $0.064 \pm 0.006$ | 0 |
| | $w_2$ | $0.005 \pm 0.004$ | 0.29 |
| | $w_3$ | $0.003 \pm 0.008$ | 0.75 |
| France | $w_0$ | $0.004 \pm 0.004$ | 0.25 |
| | $w_1$ | $0.073 \pm 0.008$ | 0 |
| | $w_2$ | $0.005 \pm 0.005$ | 0.32 |
| | $w_3$ | $-0.013 \pm 0.011$ | 0.25 |

**Table 1. Fitting the flipping probability**. Coefficients of the linear regression of Eq. 4 applied to the datasets of census tracts in US and French cities. According to the $p$-values, the coefficients $w_2$ and $w_3$ cannot be assumed to be different from 0 implying that flipping probabilities are symmetric P(-1→+1)=P(+1→-1). The timestep is here two years. We also note that in France, we can assume $w_0 = 0$.

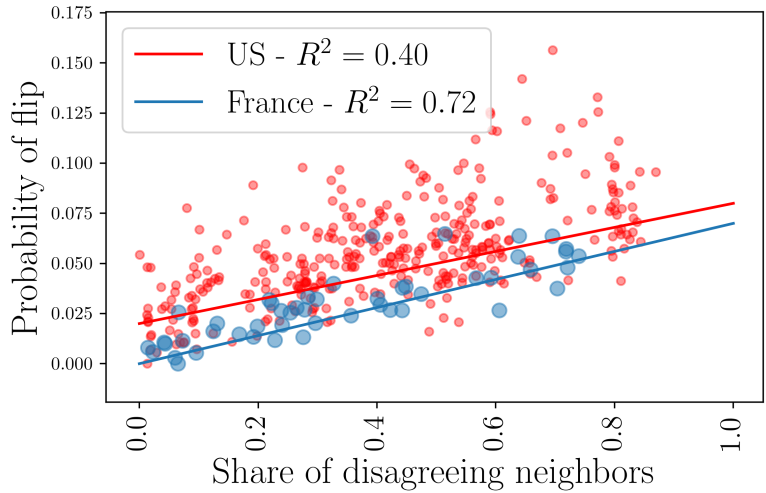

**Figure 2. Flipping probability versus the fraction of different neighbors.** Probability for a tract to flip to another income category between years $t$ and $t + \Delta t$ (with $\Delta t = 2$ years) as a function of the fraction of its direct neighbors in the opposite level. Each point represents a bin of tracts in the same city (8 bins per city). We fit the flipping probability with a linear function of the form $w_0 + w_1 f_i$ (Eq. 5). The coefficients $w_0$ and $w_1$ are found to be independent from the flipping direction.

**The voter model**

A flipping probability that follows Eq. 5 constitutes the basis of the well-known "voter model" in statistical physics (Krapivsky, et al., 2010; Liggett, 1997; Dornic, et al., 2001; Krapivsky, 1992; Frachebourg & Krapivsky, 1996) which describes consensus formation in a population of agents characterized by a set of integer (in the usual case, the state of each agent is described by a binary variable $S = \pm 1$), and can also be seen as a model for the dynamics of spatial conflict (Clifford & Sudbury, 1973). The voter model is usually defined on a graph $G = (V, E)$ and at each time step, each vertex flips at a rate $w_i(S)$ that equals the fraction $f_i$ of its neighbors in the opposite state and which reads in terms of the binary variables

$$w_i(S) = \frac{1}{2}\left(1 - \frac{S_i}{|N(i)|}\sum_{j \in N(i)} S_j\right) \quad (6)$$

In our case of urban dynamics, we consider a slight variation of the voter model on a graph $G$ where the transition rate is given by $w_i(S) = w_0 + w_1/2(1 - S_i/|N(i)|\sum_j S_j)$. In such a "noisy voter model" (Granovsky & Madras, 1995; Carro, et al., 2016), the probability of flipping is non-zero even when all agents are in the same state $S$, and in contrast with the usual voter model (Krapivsky, et al., 2010), the average total state $M = 1/|V|\sum_i <S_i>$ converges towards 0 and no consensus is reached whatever the initial conditions. In other words, in this case we always end up at large times in a situation where 50% of the agents are in the state $S = +1$ and the other half in the state $S = -1$ (see Methods.).

**Dynamics of household income**

In the following, we want to understand whether a voter model can account for the dynamics of household income in cities. In order to get some intuition about the structure of tracts in France and in the US, we measure their degree distribution (see Fig. 3). We observe that they are not far from a regular lattice of degree 6 in the US case and 5 for French cities (here, a regular lattice means that all the tracts have the same number of neighbors). In our simulations, we use the real structure of tracts and thus include possible effects due to the small degree fluctuations.

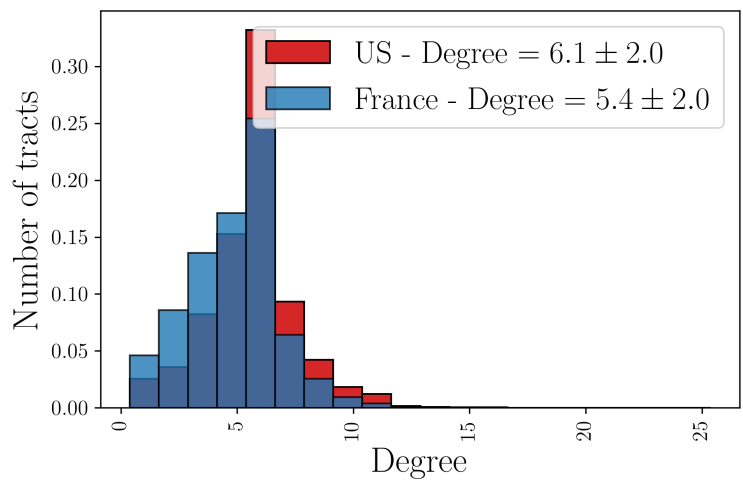

**Figure 3.** Distribution of the degree of tracts in American and French FUAs. The average degree is 6.1 with a standard deviation of 2.0. One can assume that the network of tracts in American cities is a regular lattice of degree 6 (and 5 for French cities).

At an aggregate level (for all cities and all years of the country), we have shown that the flipping probability can be described by a simple linear function of the fraction of neighbors in the opposite income level (Eq. 5). One can also fit such a relation at the city level, thus giving different $w_0$ and $w_1$ coefficients for each tract. We recover on average the coefficients resulting from a global fit with an average $w_0 = 0.03$ and $w_1 = 0.06$ for a timestep of 2 years. These two values give us two different time scales in studying the dynamics of household income in American cities. The typical flipping time for a tract to be "converted" by its neighbors is given by the time scale $t_1 \sim 1/w_1 \approx 33$ years. This timescale is much larger than the typical time length of study (10 years) which means that the interface length $l(t)$ (ie. the length of the interface between rich and poor tracts) is roughly constant at the scale of our study. In France, we find that $w_0 = 0$ and $w_1 = 0.07$ for a timestep of 2 years, which gives a model that is closer to the classic voter model and the typical time scale is of order 29 years.

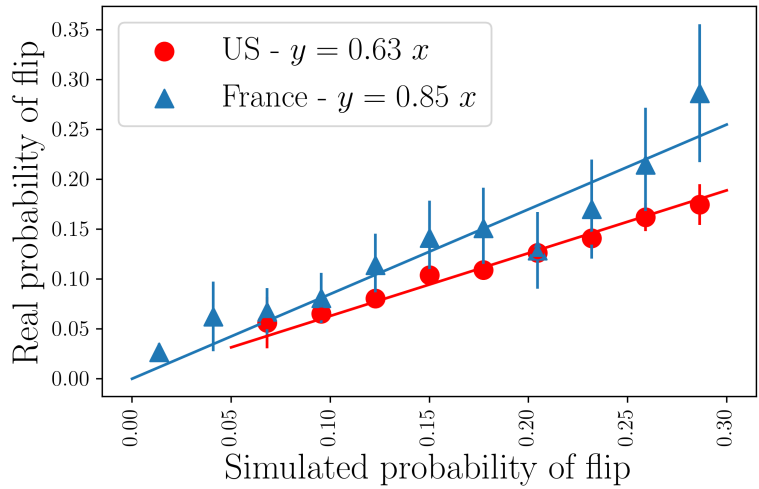

**Figure 4. Observed versus simulated flipping probabilities**. A perfect prediction corresponds to the line $y = x$ and we indeed observe here a linear positive trend between simulated and expected results for the tract flipping probabilities. However, we tend to overestimate the probability of changing level over the years (due to the existence of on-off one-year drops in the median household incomes that tend to make tracts spuriously flip and flip back to their previous status). The order in probability of flipping is however conserved: the Kendall coefficient between the two lists of probabilities is equal to 1 for the US and 0.92 for France. With only two parameters in the US and one in France, our model is able to predict which tracts are the most likely to change income level over the years.

So far, we have shown that there is a prior relation between the probability of flipping and the share of neighbors in opposite status that is akin to the evolution equation of a noisy voter model. We have however to test whether such a model is *a posteriori* able to reproduce long-term tendencies of household income in cities. We thus build a counterfactual noisy voter model with the same values for all American cities parameters $w_0 = 0.015$/year and $w_1 = 0.03$/year in the US (the parameters of Table 1 were fitted over 2 years and we have to half the values). Starting in 2010, we run 1000 simulations per city of the noisy voter model and compare each year the empirical status of tracts to the ones resulting from the simulations. After ten years of simulation, we compute the probability for every tract in American city to have flipped between 2010 and 2019. We then pool together tracts with similar predicted probability of flipping and compare the average probability of flip over the years with tracts that have actually flipped (see Fig. 4). We follow the same method for French cities, starting in 2001 and ending in 2017 with parameters $w_0 = 0$ and $w_1 = 0.03$/year (see Fig. 4). We see that there is a positive linear trend between predicted and actual probabilities of flipping for tracts in both France and the US. Yet, the coefficient of the linear regression is smaller than 1, meaning that we tend to overestimate the probability of flipping. The main reason for that is the existence of on-off one-year drops in the median household incomes that tend to make tracts spuriously flip and flip back to their previous status. Yet, Figure 4 shows that a very parsimonious model (with only 1 or 2 parameter) based on a voter model is sufficient enough to statistically predict the evolution of median household income levels at the tract scale in a city over 10-year periods. This result shows that a voter model is not only a prior descriptive tool of income dynamics in cities but can also be used in a predictive manner.

**Discussion and conclusion**

So far, we have seen that in American and French major cities, the probability for a tract to change income category strongly depends on the levels of its immediate neighbors in a voter model like relation of the form $\text{Prob}(S_i(t) \neq S_i(t-1)) \approx w_0 + w_1 f_i$ and this probability is symmetric, meaning that the values of $w_0$ and $w_1$ are the same in both directions of flips. The income evolution of an area depends mostly on the social status of its immediate neighbors (contribution of $w_1$) added with some noise (contribution of $w_0$), the contribution of neighbors being 3 times larger. We thus can build a simple and parsimonious model of income dynamics with only two parameters at the US scale and one for France. From this model, we can predict the probability for a tract to change income category over 10-year periods of time (see Fig. 4). If our estimation is not perfect and tends to overestimate the probability of changing level, the result is surprisingly encouraging for a 2-parameter model. Moreover, if the probability of flipping is overestimated, we can see on Fig. 4 that it can be rectified by a simple linear factor, meaning in particular that the order in probability of flipping is conserved.

Although this model is not perfect it could serve a starting point for more refined analysis and provides a very simple tool for exploring various evolution scenarios for urban areas in the future. In order to illustrate this point, we show on Fig. 5 a prediction of the income evolution of Boston tracts and Paris IRIS up to 2030. Starting in 2019, we simulate 1000 evolutions of Boston and Paris tract levels with the voter model previously described. For each tract, we compute the probability to flip between 2019 and 2030. The darker zones correspond to higher probabilities of flipping. Blue zones are below-average income areas for which we compute the probability to get above average. Red areas are above-average income areas for which we compute the probability to drop below average. The most interesting tracts are the dark ones, either blue or red, in the sense that there is a large probability to see them flipping in the next decade. This predictive result shows that a voter model applied to income dynamics is not only descriptive but can capture more fundamental elements of city dynamics.

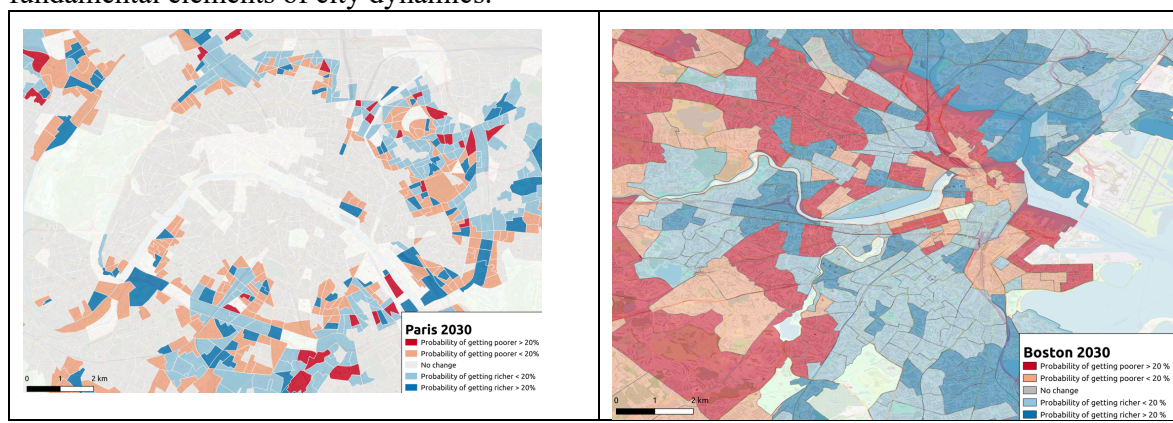

**Figure 5. Predicted probability of changing income category for Paris and Boston**. We simulate 1000 evolutions of Boston and Paris tracts with the voter model described in the text. For each tract, we compute a probability to flip between 2019 and 2030 for Paris (left) and Boston (Right). Darker tracts are the tracts with highest probabilities of flipping. Red areas are above-average income areas for which we compute the probability to drop below average. Most interesting tracts are the dark ones, either blue or red, that have a large probability to flip in the next decade.

In this paper, we investigated the change of income status of a specific urban area. We showed that the change of the income status of a specific neighborhood can be described by a parsimonious model that gives both descriptive and predictive insights for urban analysis. In a broad sense, the process of getting gentrified or impoverished proxied here by a probability to switch from one income status to another is well approximated by a voter model, meaning in particular that the effects of immediate neighbors are essential in the socio-economic dynamics of a district. We used here a binary representation of income levels in the city and a continuous income approach could lead to a more refined description of these processes. This would require to write a spatiotemporal differential equation for the income state and efforts in this direction constitute an interesting topic for future research.

**Data and materials availability:** All the data used in this paper is freely available for download at the URL given in references (Manson, et al., 2021) and (INSEE, 2022). The code written in Python is available upon request.

# Methods

## Datasets

In the US, our analysis is based on the available data of household incomes in American census tracts for each year between 2010 and 2019 (Manson, et al., 2021). This data is provided by the IPUMS-NHGIS (for the National Historical Geographic Information System) and comprises population, housing, agricultural, and economic data, along with GIS-compatible boundary files for geographic units in the United States from 1790 to the present. In particular, this dataset contains time series tables that link together comparable statistics from multiple censuses using standardized categories and codes.

We cross these data with the delineations of 159 OECD Function Urban Areas (FUA) in the US to produce a dataset of 159 cities with household income at the census tract level, of which we keep cities with more than 200 tracts, which leaves us with 45 cities.

In France, our analysis is based on the available data of household incomes in small areas called IRIS (which stands for "Ilots Regroupés pour l'Information Statistique") for each year between 2001 and 2017 (INSEE, 2022). We cross these data with the delineations of 85 FUAs in France to produce a dataset of 85 cities with household income at the census tract level, of which we keep cities with more than 150 IRIS, which leaves us with 9 cities.

**Data smoothing**

Since we observe unusual one-year drops of household incomes in the data, we use a convolution regularization to smooth out one-year outliers. Indeed, some tracts exhibit unexplained one-year drops or increases of average income that make them change income level for just a year without demonstrating a long-time trend. In order to smooth out such problems, we decided to do a double-step check of income levels. We first compute the level of each tract every year, then recompute the level of each year as the average level over three years via a majority rule:

$$S_i(t) = \text{Round}\left(\frac{1}{3}(S_i(t-1) + S_i(t) + S_i(t+1))\right)$$

**Time step**

To compare the probability of changing level with the share of neighbors in the opposite level, we compare the level of each tract $s_i(t)$ at each time step to its previous level $s_i(t-1)$ and the previous levels of the neighbors. Yet, we use in our code a timestep of two years in order to smooth out noisy behaviors and single out long-term trends.

**Average total state**

The state $<S_i>$ at site $i$ averaged over all possible configurations evolves as (Krapivsky, et al., 2010)

$$\frac{d<S_i>}{dt} = -2<S_i w_i> = -<S_i> + \frac{1}{|N(i)|}\sum_{j \in N(i)} <S_j> \quad (7)$$

If the graph $G$ is a regular lattice with $|N(i)| = z$, using this equation it is not difficult to show that the average total status $M = 1/|V|\sum_i <S_i>$ is conserved. As a result, if $\rho$ denotes the initial fraction of +1 states in $G$, the probability of reaching a final +1 consensus is given by $\rho$. In our case of urban dynamics, we consider a slight variation of the voter model on a graph $G$ where the transition rate is given by $w_i(S) = w_0 + w_1/2(1 - S_i/|N(i)|\sum_j S_j)$. In such a "noisy voter model" [31, 32], the probability of flipping is non-zero even when all sites are in the same state.

We observe that the degree distribution of tracts within American and French FUAs can be well approximated by a regular lattice of degree 6 (see Fig. 3) and if we consider the voter model on a hypercubic lattice ($|N(i)| = 2d$), we then obtain using the equation for the time evolution of the average state $<S_i>$ at site $i$, that the average total status $M$ evolves as

$$M(t) = M(0)e^{-2w_0 t} \quad (8)$$

Hence, this noisy voter model converges on average towards a null state ($M = 0$) and no consensus is reached whatever the initial conditions. This due to the effect of noise introduced by the intercept $w_0$: even when all sites are in the same state, there are on average $w_0 N \Delta t$ sites that flip during time $\Delta t$. We also recover the fact that when $w_0 = 0$, the total average state is constant as expected for the classical voter model (Krapivsky, et al., 2010).